\documentstyle[epj,epsfig,final]{svjour}

\begin{document} 
\headnote{Rapid Note}

\title{Mott transition of the $f$-electron system in the periodic Anderson model with nearest neighbor hybridization}
\author{K. Held\thanks{\email{held@physik.uni-augsburg.de}} 
and R. Bulla}
\authorrunning{K.~Held and R. Bulla}
\titlerunning{Mott transition of the $f$-electron system in the PAM}
\institute{Theoretische Physik III, Elektronische Korrelationen und Magnetismus, Universit\"at Augsburg,
86135 Augsburg, Germany\\
}
\date{\today}

\abstract{
We show analytically that, under certain assumptions, 
the
periodic Anderson model and the Hubbard model
become equivalent  within the
dynamical mean field theory for quasiparticle weight $Z\rightarrow 0$.
A scaling relation is derived which is validated numerically
using  the numerical renormalization group  at zero temperature
and  quantum Monte Carlo simulations at finite temperatures.
Our results show that the $f$-electrons of the 
half-filled 
periodic Anderson model with nearest neighbor hybridization
get localized at a finite critical interaction strength $U_{\rm c}$,
also at zero temperature. 
This transition is equivalent to
the Mott-transition in the Hubbard model.}
\PACS{{71.27.+a} {Strongly correlated electron systems; heavy fermions}
\and{71.10.Fd}{Lattice fermion models (Hubbard model, etc.)}\and
{71.30.+h}{Metal-insulator transitions and other electronic transitions}
}
\maketitle

The periodic Anderson model (PAM) and the Hubbard model (HM) are two of the
most fundamental models in condensed matter physics.
Despite the simplicity of their  Hamiltonians, the many body nature of these models results
in a complicated correlated electron problem which does not allow for an exact solution except for the one-dimensional HM.
Also in infinite dimensions \cite{DMFT,Georges}, where both models map
onto a single impurity Anderson model with different self-consistency conditions, an exact solution is
only possible numerically. In a recent paper \cite{Held99}, such a numerical study
showed an astonishingly similar behavior of the two models and, in particular, that the PAM exhibits a 
transition similar to the Mott transition of the HM.

In the present paper we show analytically
that under certain  assumptions
the infinite dimensional PAM and the HM become equivalent in the limit 
of vanishing quasiparticle weight 
$Z$. These assumptions are similar in nature to those
employed for the self-consistent projective method
\cite{Moeller} and the linearized dynamical mean-field theory \cite{BullaPotthof}.
Thus, the hitherto  numerically found similarity can be understood and the
critical Coulomb interaction $U_c$ for the Mott transition
of one model can be determined from that of the other model.
Applying the numerical renormalization group (NRG) at zero temperature ($T=0$)
and the quantum Monte Carlo (QMC)
technique at finite $T$, we investigate to what extent the two underlying
assumptions are fulfilled
and show, for the first time, that  the zero temperature
 PAM with nearest neighbor hybridization
has a Mott transition with quasiparticle weight
$Z\rightarrow 0$
at a finite  $U_c$, in contrast to the single impurity Anderson model
for which  $Z\rightarrow 0$ at $U_c=\infty$.

The Hamiltonian of the HM reads
\begin{equation}
H = 
- t \sum_{ \langle i j \rangle \sigma}
f_{i \sigma}^{\dagger} f_{j \sigma}^{\phantom{\dagger}} 
+ U \sum_{i} (n_{i\uparrow}^{{f}} - \frac12)
(n_{i\downarrow}^{{f}} - \frac12).
\label{hmham}
\end{equation}   
Here, $f_{i \sigma}^{\dagger}$ and  $f_{j \sigma}^{\phantom{\dagger}}$ are
creation and annihilation operators for an electron with spin $\sigma$ on site $i$ or $j$, respectively,
$n_{i\sigma}^{{f}}= f_{i \sigma}^{\dagger}f_{i \sigma}^{\phantom{\dagger}}$, $\langle i j \rangle$ denotes 
the sum over
nearest neighbors, and $t$ the hopping amplitude between them.
We use the symbol $f$ for the electrons of the HM since the equivalence of these
and  the $f$-electrons of the PAM will be reported in this paper.
The PAM consists of a band of conducting electrons 
($d$-electrons) and interacting $f$-electrons.
Both are coupled via the hybridization $V_{ij}$:
\begin{eqnarray}
  H &=& - t \sum_{ \langle i j \rangle \sigma}
d_{i \sigma}^{\dagger} d_{j \sigma}^{\phantom{\dagger}}
  +  \sum_{ i j \sigma} V_{i j}\;
  (d_{i\sigma}^{\dagger} f_{j\sigma}^{\phantom{\dagger}}
  + h.c.) \nonumber \\ && + U \sum_{i} (n_{i\uparrow}^{{f}} - \frac12)
  (n_{i\downarrow}^{{f}} - \frac12).
\label{pamham}
\end{eqnarray}   
We only consider the particle-hole symmetric case ($\mu=0$ in this form for
 a symmetric non-interacting density of states (DOS).
 In  infinite dimensions or with the number of nearest neighbors
 ${\cal Z} \rightarrow \infty$, 
a non-trivial scaling of the kinetic energy
is obtained by $t=t^*/\sqrt{{\cal Z}}$. In the following, $t^*\equiv 1$ sets the
energy scale.
We consider two different kinds of hybridizations: (i) a nearest neighbor hybridization
$V_{i j}= t_{df}/\sqrt{{\cal Z}}$ for nearest neighbors $i$ and $j$ which is
 zero otherwise \cite{PAMi} and (ii) an
on site  hybridization with $V_{i i}= t_{df}$ and zero otherwise \cite{PAMii}.

Within dynamical mean field theory (DMFT) \cite{DMFT,Georges}, 
which becomes exact in infinite dimensions,
the HM and the PAM map onto the same single site problem (which depends on the Green function $G_f$, 
self-energy $\Sigma_f$, $T$, and $U$) but different self-consistency conditions.
For the HM this  self-consistency condition at frequency $\omega$ is given by
\begin{eqnarray}
G_f(\omega) &=& \int d\epsilon \; \frac{N(\epsilon)}{\omega-\Sigma_f(\omega)-\epsilon}, 
\end{eqnarray}
where $N(\epsilon)$ is the non-interacting DOS.
In the case of the PAM, the $d$-electrons can be integrated out since they enter only
quadratically in the Hamiltonian and the effective action. This results in an effective $f$-electron
problem with  a self-consistency condition that reads
\begin{eqnarray}
 G_f(\omega)= \int d\epsilon\; \frac{N(\epsilon)}{\omega-\Sigma_f(\omega)-t_{df}^2\epsilon^2/(\omega-\epsilon)}
\label{nn}
\end{eqnarray}
for the PAM with nearest neighbor hybridization and
\begin{eqnarray}
G_f(\omega)= \int d\epsilon\; \frac
{N(\epsilon)}{\omega-\Sigma_f(\omega)-t_{df}^2/(\omega-\epsilon)}\label{onsite}
\end{eqnarray}
for the PAM with on site hybridization, where
$N(\epsilon)$ is the free $d$-electron DOS. Note, that the effective
one-particle potential of the PAM ($\propto 1/(\omega-\epsilon)$)
is 
frequency dependent, i.e., retarded, due to the fact that the electrons may move from the $f$-orbitals
to the $d$-band and return at a later time.
The main difference between nearest neighbor [Eq.(\ref{nn})] and on site 
hybridization  [Eq.(\ref{onsite})] is that the former describes metallic
f-electrons at $U=0$ and within a Fermi liquid phase while the latter
describes Kondo-insulating f-electrons, i.e., a gapped 
 f-electron quasiparticle  peak induced by the  hybridization.

The equivalence of the PAM and the HM (at the Mott transition $Z\rightarrow 0$)
is shown on the basis of two assumptions: (i) that the metallic phase of the $f$-electrons
may be described by Fermi liquid theory at low energies and (ii)
that the remaining spectral weight of $1-Z$ is contained in two
 Hubbard bands centered around $\pm U/2$ and,
in particular, that differences in the internal structure of these  bands
have no influence on the low-energy physics.
Assumption (ii) is certainly only fulfilled approximately and
becomes justified if the high energy features are well separated from the 
low-energy features \cite{Noteii}. For a detailed discussion on this assumption
see Sec. 2 of \cite{BullaPotthof}.

With  assumption (i) and $Z=(1-\partial \Sigma_f/\partial \omega|_{\omega=0})^{-1}$
the low-frequency self-consistency condition for the PAM with
nearest neighbor hybridization is given by
\begin{eqnarray}
   G_f(\omega)\! &\!=\!&\! 
   \int\! d\epsilon  \frac{Z N(\epsilon)}
   {\omega-Z t_{df}^2\epsilon^2/(\omega-\epsilon)}
 \\   &=& 
   \int \! d\epsilon \frac{Z N(\epsilon)}
   {\omega-\epsilon/2+ \sqrt{1+4Zt_{df}^2}\epsilon/2}\frac{\sqrt{1\!+\!4Zt_{df}^2}\!+\!1}{2\sqrt{1\!+\!4Zt_{df}^2}}
   \nonumber\\&&+
    \frac{Z N(\epsilon)}{\omega-\epsilon/2- \sqrt{1+4Zt_{df}^2}\epsilon/2}\frac{
 \sqrt{1\!+\!4Zt_{df}^2}\!-\!1}{2\sqrt{1\!+\!4Zt_{df}^2}}.
\label{partfrac2}
\end{eqnarray}
With the partial fraction decomposition above and a variable transformation 
$y=\frac{1}{2Z}(1\mp\sqrt{1+4Zt_{df}^2}) \epsilon$ 
for the two terms of Eqn. (\ref{partfrac2}) one obtains
\begin{eqnarray}
 G_f(\omega)&=& \int  dy\; \frac{Z\tilde{N}(y)}{\omega-Zy}\label{PAMHM}, 
\end{eqnarray}
which is just the low-frequency self-consistency condition of the HM
within a Fermi liquid phase. However, 
with a DOS which depends on $Z$ (which is itself a function of $U$):
\begin{eqnarray}
\tilde{N}(y) \!&\!&\!=\! N\!\Bigg(\!{\scriptsize\frac{2Zy}{\sqrt{1\!\!+\!4Zt_{\!df}^2}\!-\!\!1}}\!\Bigg)
\frac{\sqrt{1\!+\!4Zt_{df}^2}\!+\!1}{\sqrt{1\!+\!4Zt_{df}^2}\!-\!1}\frac{Z}
{\sqrt{1\!+\!4Zt_{df}^2}}
\nonumber\\&&\!+\!
N\!\Bigg(\!\frac{2Zy}{\sqrt{1\!\!+\!4Zt_{\!df}^2}\!+\!\!1}\!\Bigg)
\frac{ \sqrt{1\!+\!4Zt_{df}^2}\!-\!1}{\sqrt{1\!+\!4Zt_{df}^2}\!+\!1}\frac{Z}
{\sqrt{1\!+\!4Zt_{df}^2}}.
\end{eqnarray}
In the limit $Z\rightarrow 0$,  $\tilde{N}(y)$ reduces to
\begin{eqnarray} 
\tilde{N}(y) &\stackrel{Z\rightarrow 0}{\longrightarrow}& 
 N(y/t_{df}^2)/t_{df}^2 +{\cal O}(Z^2).\label{Escale}
\end{eqnarray}
and becomes, thus, independent of $Z$. Therefore, at  $Z\rightarrow 0$, 
the low energy spectral function
of the PAM is identical to that of the HM with the DOS of Eq.~(\ref{Escale}).
With assumption (ii), i.e., that differences in the internal structure of 
the Hubbard bands 
have no impact on the low-energy physics, the PAM with nearest neighbor
hybridization is equivalent to a HM which has the DOS of the PAM's $d$-electrons
renormalized by the factor $1/t_{df}^2$. Thus, the critical Coulomb interaction and temperature
for the Mott transition of the PAM can be calculated from that of the HM via
\begin{eqnarray}
 U_c^{{\rm{PAM}}}=  t_{df}^2\; U_c^{{\rm{HM}}} &\phantom{aa}{\rm{and}}\phantom{aa}& T_c^{{\rm{PAM}}}=t_{df}^2\; T_c^{{\rm{HM}}}  .  \label{Escale2}
\end{eqnarray}
For the PAM with on site hybridization, the same proceeding yields for
$Z\rightarrow 0$:
\begin{eqnarray}
 G_f(\omega)  &\stackrel{Z\rightarrow 0}{\longrightarrow}&   \int_{-\infty}^{\infty} dy\;  \frac{Z N(-t_{df}^2/y)}
  {\omega-Zy} \frac{t_{df}^2}{y^2}
\end{eqnarray}
Again, this is equivalent to the low-frequency self-con\-sis\-tency condition of a HM.
While the PAM with nearest neighbor hybridization 
(which has metallic $f$-electrons at small $U$)
 maps to a familiar HM, the PAM
with on site hybridization
(which yields a Kondo insulating  $f$-electron system)
 maps to a rather unusual HM.
If the free $d$-electron DOS of the PAM $N(\epsilon$) is zero for $|\epsilon|>D$, the 
PAM with on site hybridization
is equivalent to a HM with a gap of size $t_{df}^2/D$ in the non-interacting DOS. This reflects the Kondo insulating nature of this model.
Furthermore, the DOS of this unusual HM has not a finite bandwidth 
but tails decaying like $1/y^2$ for large energies. In particular, the 
 standard deviation of this DOS is infinite. For such a DOS, the linearized DMFT \cite{BullaPotthof} predicts a Mott transition at $U_c=\infty$ and for the
HM with Lorentzian DOS (a similar DOS without gap) it is known that 
 $U_c=\infty$ \cite{Georges92}. Thus, at $T=0$ 
the analytic argument suggests  $U_c=\infty$
for the PAM with on site hybridization,
in agreement with recent NRG results \cite{PruschkePC}. Nevertheless, at finite
$T$ the transition is very similar to that of the HM and the PAM with
nearest neighbor hybridization \cite{Held99}. This is due to the fact
that the vanishing of the quasiparticle peak at a fixed finite
temperature is unaffected by the very small energy scale
present at $T=0$.

\begin{figure}
\unitlength1cm

{\psfig{figure=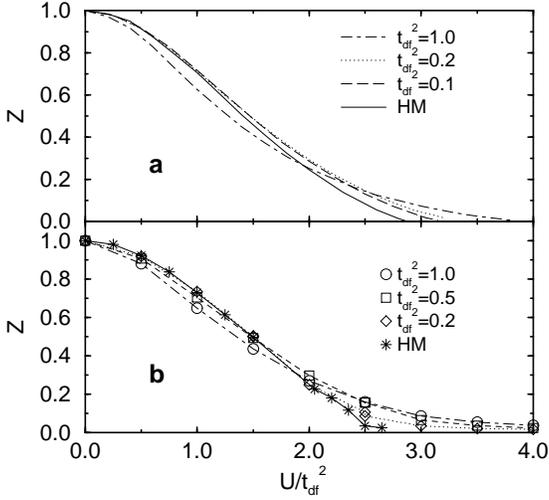,width=8.cm,angle=0}}

\caption{Quasiparticle weight $Z$ 
as a function of $U/t_{df}^2$ for the PAM with nearest neighbor 
hybridization at a) $T=0$ (NRG) and b) $T/t_{df}^2=0.05$ (QMC)
in comparison to that of the HM. The QMC data for the HM are partly 
from  \protect \cite{Schlipf99a}.
\label{ZvsU}}
\end{figure}

With the approximative nature of assumption (ii) in mind we 
investigate now numerically by NRG \cite{NRG} and QMC \cite{QMC} 
to what extent the derived scaling relations 
hold. To this end, we calculate $Z$ \cite{calcZ} 
as a function of $U/t_{df}^2$ for a Bethe lattice. The results
are presented in 
Fig.~\ref{ZvsU} which shows that
there is  indeed a Mott
transition $Z\rightarrow 0$ at $T=0$ in the PAM
with nearest neighbor hybridization. As in the HM \cite{Georges,Bulla99a,Roz99},  the 
coexistence of two solutions is observed at  $T=0$ (Fig.~\ref{ZvsU} only contains
the metallic solution obtained with increasing $U$).
Fig.~\ref{ZvsU}
validates, furthermore, that the scaling relation
Eqn.~(\ref{Escale2}) holds, at least approximately,
even though, the actual values of $U$
differ by a factor of 10, both at zero and finite temperature
(at finite temperature, $T/t_{df}^2$ was kept constant
according to Eq.(11)).

\begin{figure}
\unitlength1cm

\begin{center}
{\psfig{figure=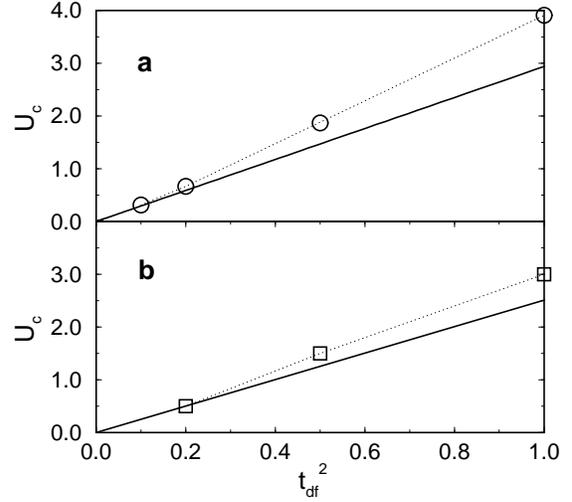,width=8cm,angle=0}}
\end{center}

\caption{Critical value $U_c$ for the Mott transition
of  the PAM with nearest neighbor 
hybridization at a) $T=0$ (circles) and b) $T/t_{df}^2=0.05$ (squares)
compared to $t_{df}^2 U_c^{{\rm{HM}}}$ (solid line) at these temperatures. 
\label{Uvst}}
\end{figure}

From Fig.~\ref{ZvsU} the critical value of $U$ for the Mott transition,
i.e., the vanishing of $Z$,
is determined \cite{calcU}.
The results are compared to  prediction (\ref{Escale2}) in Fig.~\ref{Uvst}.
At small $t_{df}$, 
the agreement is very good, while there are notable deviations 
at larger $t_{df}$. These deviations can be understood from 
the spectral functions discussed below.

\begin{figure}
\unitlength1cm

\hspace{-.325cm}{\psfig{figure=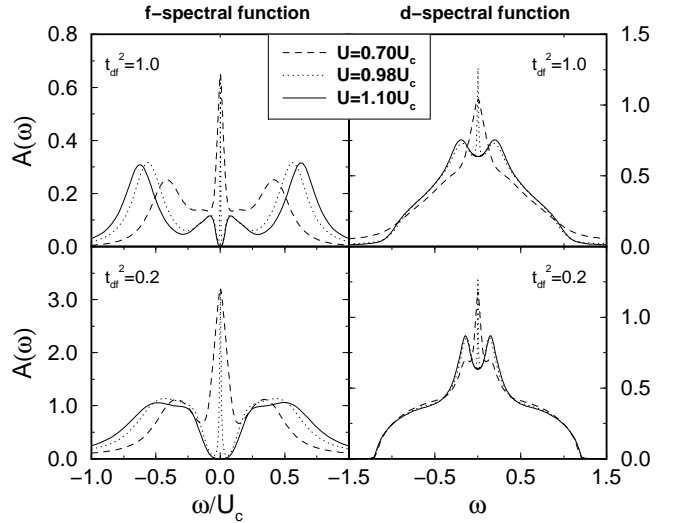,width=9.2cm,angle=0}}
\caption{Spectral function of the $f$- and $d$-electrons for the
PAM with nearest neighbor 
hybridization 
at $T=0$ (NRG) and at different values of $U$ close to
the Mott transition. \label{spectrum}}
\end{figure}

Fig. \ref{spectrum} shows the disappearance of the central
quasiparticle peak at the Mott transition 
in the $f$-electron spectrum of the
PAM with nearest neighbor 
hybridization. At the same time, the $d$-electron spectral function
 remains finite at the Fermi energy. Thus, despite
the Mott transition  of the $f$-electrons the overall 
system remains metallic. W.r.t. the  deviations between $U_c$
and prediction (\ref{Escale2}), one observes 
that at $t_{df}^2=0.2$ the quasiparticle
resonance is well separated from the high-energy Hubbard bands,
while at  $t_{df}=1$ there is additional spectral weight very close to the
quasiparticle resonance.
Thus, assumption (ii) is not a good approximation for larger $t_{df}$
with the consequence that the analytic calculation based on assumption (ii)
is less justified and  prediction (\ref{Escale2})
less accurate. This explains the the $t_{df}$-dependence of the
deviations in Fig.2 which is a priori not
clear from the analytic calculation.

Finally, Fig. 4 shows a comparison between the $f$-spectral functions for
the periodic Anderson model with nearest neighbour 
hybridization and the Hubbard model. The density of states for
the Hubbard model calculation is chosen according to
Eq. (10). As expected,
the results show a good agreement in the low-frequency 
part whereas the deviations are more pronounced in the 
high-frequency regime. The agreement for small $\omega$ is, however, 
not perfect, even in the limit $Z\to 0$, as the derivation of Eq. (10)
is only approximate.

\begin{figure}
\unitlength1cm

\begin{center}
{\psfig{figure=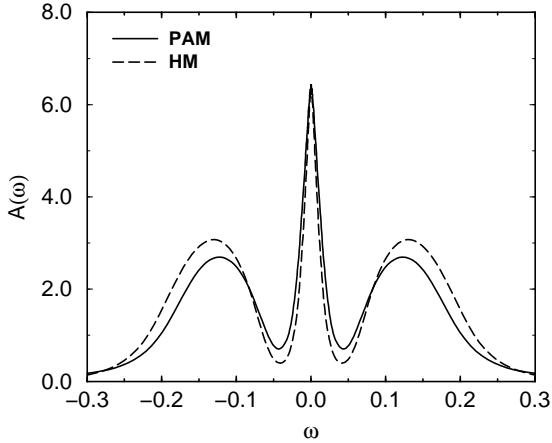,width=8cm,angle=0}}
\end{center}
\caption{$f$-spectral functions of
the periodic Anderson model with nearest neighbour 
hybridization compared to the corresponding Hubbard model according
to Eq. (10) at
$U=0.25$, $t_{df}^2=0.1$, and $T=0$ (NRG).}
\end{figure}

In conclusion, we showed analytically that the PAM 
becomes equivalent to the HM at the Mott transition $Z\rightarrow0$
if (i) the low energy physics of both models
is described by Fermi liquid theory and
(ii) the high energy features are Hubbard bands which are
well separated from the low energy quasiparticle peak.
This allows to calculate the critical interaction $U_c$ at
which  $Z\rightarrow0$ for the PAM
from that of the Hubbard model.
In particular, the PAM with on-site hybridization maps to a Hubbard model
with gapped DOS and Lorentzian tails which suggests $U_c=\infty$,
while the PAM with nearest neighbor hybridization maps
to the Hubbard model with the same DOS as the $d$-electron DOS of the PAM.
The latter leads to the scaling 
relation $U_c^{{\rm{PAM}}}= t_{df}^2 U_c^{{\rm{HM}}}$. 
Numerical calculations employing NRG and QMC yield that 
the PAM with nearest neighbor
hybridization has indeed a Mott transition at a finite $U_c$
 and that the above
scaling relation is correct for not too large values of $t_{df}$.
The similarity between both models includes the
existence of hysteresis for the PAM with nearest neighbor hybridization.

{\small We are grateful to N. Bl\"umer,
C. Huscroft, A.K. McMahan, Th. Pruschke, R.T. Scalettar, and D. Vollhardt 
for valuable discussions.
This work was supported in part by the Sonderforschungsbereich 484 of the 
Deutsche Forschungsgemeinschaft.
}


\begin{thebibliography}{10}
\bibitem{Noteii}
For the HM, the QMC simulation \cite{Schlipf99a} 
shows  that besides the 
central resonance there is additional spectral weight at low frequencies,
also close to the Mott transition. Thus, at higher temperatures, 
 approximation (ii) is less good
than at  $T=0$ where 
NRG results \cite{Bulla99a} show a more complete separation
of the central resonance from the rest of the spectrum.
\bibitem{calcZ}
To calculate $Z$ within NRG, $Z=(1\!-\!\partial {\rm{Re}}\Sigma(\omega)/\partial \omega|_{\omega=0})^{-1}$
is directly employed.
Within QMC, $Z$ is determined via the finite differential quotient
$(1-{\rm{Im}}\Sigma(i\omega_0)/\omega_0)^{-1}$ at the lowest Matsubara frequency $\omega_0$.
The latter gives a small $Z$ at finite temperatures, even after the disappearance of
the quasiparticle resonance. The QMC time discretization $\Delta \tau$ was
$\Delta \tau=0.25$, 0.333, and 0.5 for $t_{df}^2=1,$ 0.5, and 0.2, respectively.
\bibitem{calcU}
To determine $U_c$ within QMC, we decide on metal or insulator via the approximate criterion.
${\rm{Im}}\Sigma(\omega_0)< {\rm{Im}}\Sigma(\omega_1)$: indicates $\Sigma(\omega)\stackrel{\omega\rightarrow 0}{\longrightarrow}0$, i.e, a metal;
${\rm{Im}}\Sigma(\omega_0)< {\rm{Im}}\Sigma(\omega_1)$: insulator.
\end{thebibliography}
\end{document}